\documentclass[prl,superscriptaddress,preprintnumbers,tightenlines,nofootinbib]{revtex4-2}

\usepackage{amsmath}
\usepackage{amsfonts}
\usepackage{amssymb}
\usepackage{bm}
\usepackage{mathrsfs}
\usepackage{graphicx}
\usepackage[colorlinks]{hyperref}
\usepackage[usenames]{color}
\definecolor{darkgreen}{rgb}{0,0.5,0}
\hypersetup{urlcolor=darkgreen}
\usepackage{empheq}
\usepackage{ulem}
\hypersetup{
 unicode=false,          
 pdftoolbar=true,        
 pdfmenubar=true,        
 pdffitwindow=false,     
 pdfstartview={FitH},    
 pdftitle={My title},    
 pdfauthor={Author},     
 pdfsubject={Subject},   
 pdfcreator={Creator},   
 pdfproducer={Producer}, 
 pdfkeywords={keyword1} {key2} {key3}, 
 pdfnewwindow=true,      
 colorlinks=true,       
 linkcolor=red,          
 citecolor=cyan,        
 filecolor=magenta,      
 urlcolor=darkgreen,           
 linktocpage=true
}

\allowdisplaybreaks

\DeclareSymbolFontAlphabet{\mathrsfs}{rsfs}
\DeclareMathAlphabet{\mathcal}{OMS}{cmsy}{m}{n}

\newcommand{\dd}{\mathrm{d}}
\newcommand{\di}{\mathrm{i}} 
 
\newcommand{\dM}{\mathrm{M}}
\newcommand{\dm}{\mathrm{m}}

\makeatletter
\g@addto@macro\bfseries{\boldmath}
\makeatother

\begin{document}
	
\title{Gravitational-Wave Phasing of Quasi-Circular Compact Binary Systems \\ to the Fourth-and-a-Half post-Newtonian Order}

\author{Luc \textsc{Blanchet}}\email{luc.blanchet@iap.fr}
\affiliation{$\mathcal{G}\mathbb{R}\varepsilon{\mathbb{C}}\mathcal{O}$, 
	Institut d'Astrophysique de Paris,\\ UMR 7095, CNRS, Sorbonne Universit{\'e},
	98\textsuperscript{bis} boulevard Arago, 75014 Paris, France}

\author{Guillaume \textsc{Faye}}\email{faye@iap.fr}
\affiliation{$\mathcal{G}\mathbb{R}\varepsilon{\mathbb{C}}\mathcal{O}$, 
	Institut d'Astrophysique de Paris,\\ UMR 7095, CNRS, Sorbonne Universit{\'e},
	98\textsuperscript{bis} boulevard Arago, 75014 Paris, France}
\affiliation{Centre for Strings, Gravitation and Cosmology, Department of Physics,
Indian Institute of Technology Madras, Chennai 600036, India}

\author{Quentin \textsc{Henry}}\email{quentin.henry@aei.mpg.de}
\affiliation{Max Planck Institute for Gravitational Physics\\
	(Albert Einstein Institute), D-14476 Potsdam, Germany}

\author{Fran\c{c}ois \textsc{Larrouturou}}\email{francois.larrouturou@desy.de}
\affiliation{Deutsches Elektronen-Synchrotron DESY,\\
Notkestr. 85, 22607 Hamburg, Germany}

\author{David \textsc{Trestini}}\email{david.trestini@obspm.fr}
\affiliation{Laboratoire Univers et Théories, Observatoire de Paris, Université PSL, Universit\'e Paris Cit\'e, CNRS, F-92190 Meudon, France}
\affiliation{$\mathcal{G}\mathbb{R}\varepsilon{\mathbb{C}}\mathcal{O}$, 
	Institut d'Astrophysique de Paris,\\ UMR 7095, CNRS, Sorbonne Universit{\'e},
	98\textsuperscript{bis} boulevard Arago, 75014 Paris, France}

\date{\today}

\begin{abstract}
The inspiral phase of gravitational waves emitted by spinless compact binary systems is derived through the fourth-and-a-half post-Newtonian (4.5PN) order beyond quadrupole radiation, and the leading amplitude mode $(\ell,\dm)=(2,2)$ is obtained at 4PN order. We also provide the radiated flux, as well as the phase in the stationary phase approximation. Rough numerical estimates for the contribution of each PN order are provided for typical systems observed by current and future gravitational wave detectors.
\end{abstract}


\preprint{DESY-23-043}

\maketitle

At the time when the LIGO and Virgo gravitational-wave detectors were approved, there was no theoretical prediction available for gravitational waves (GWs) generated by compact binary systems, apart from that of the famous Einstein quadrupole formula~\cite{E18,LL,PM63}. However, it was soon realized that, given the frequency band and the expected sensitivity of these ground-based detectors, the waveform modeling was to be drastically improved in order to extract all the potential information from the signal, at least in the case of the inspiral of two neutron stars~\cite{3mn,CF94}. The breakthrough came with the merging of the post-Newtonian (PN) and the multipolar post-Minkowskian (MPM) expansions into a single formalism~\cite{BD86,B87,BD88,BD92,B98mult}, that was applied with success to derive step by step the waveform of compact binary systems up to 3.5PN order~\cite{BDI95,BDIWW95,BIWW96,B96,B98tail,BIJ02,BFIJ02,BI04mult,BDEI04,BDEI05dr} (the results are also known at 4PN order for the spin-orbit coupling and the spin-spin coupling, see \emph{e.g.}~\cite{BlanchetLR,cho2022gravitational}). Since then, many works (outlined below) have aimed at extending the precision of this result to the next level, namely 4PN or even 4.5PN order beyond the Einstein quadrupole formula.

This Letter provides the final results of these efforts, \emph{i.e.}, the GW phasing of non-spinning compact binary systems on quasi-circular orbits up to 4.5PN order, as well as the dominant GW mode, given by $(\ell,\dm)=(2,2)$, at 4PN order. Ready to be used for building accurate PN template banks for the detection and analysis of the inspiral phase of compact binaries, they should be important for third generation ground-based detectors (Einstein Telescope and Cosmic Explorer), future space-borne detectors (LISA and TianQin) and of course the current second-generation detectors (LIGO, Virgo and KAGRA). All results presented in this Letter are to be found in the ancillary file~\cite{AncFile} associated with the companion paper~\cite{BFHLT_II}.

Besides improving the detectors' data analysis, the motivation for computing high PN orders is also to perform high-accuracy tests of general relativity (GR), since the PN coefficients directly probe the non-linear structure of the theory. By confronting results from the PN expansion against data, one can put constraints on potential deviations from GR~\cite{Bsat95,AIQS06a}. This has already allowed for the confirmation of the signature of GW tails~\cite{LIGOtestGR,LIGOtestGR2}, and is promising for tests with future multi-band detections between LISA and ground-based detectors~\cite{Sesana16}.

We denote by $f(t)$ the frequency of the dominant $(2,2)$ mode of the GW as measured by an observer in the asymptotically flat region far from the source (recall that this is twice the orbital frequency), and by $\psi(t) = \pi \int \! \dd t f(t)$ the corresponding half-phase. As usual, we define the directly-measurable PN parameter $x = \mathcal{O}(c^{-2})$ by
\begin{equation}\label{xPN}
	x\equiv \left(\frac{\pi G m f}{c^3}\right)^{2/3}\,,
\end{equation}
where $m=m_1+m_2$ is the binary's total mass, $m_1$ and $m_2$ being the constant masses of the progenitors. For circular orbits, $x$ may be  defined invariantly from the Killing vector of the helical symmetry in the asymptotically flat space-time. Since  compact binaries tend to have circularized by the time they enter the detector's frequency band~\cite{Peters64}, we only consider the case of quasi-circular orbits, for which the time-evolution of the frequency and phase (or ``chirp'') is entirely driven by the energy flux-balance equation,
\begin{equation}\label{balance}
	\frac{\dd E}{\dd t} = - \mathcal{F}\,,
\end{equation}
where $E$ denotes the invariant energy of the compact binary and $\mathcal{F}$ the total energy flux (or GW luminosity). Both~$E$ and $\mathcal{F}$ in the balance equation are unique functions of the PN parameter $x$ and the two masses. They have to be evaluated with the same relative PN precision, in the present case 4.5PN ($\sim x^{9/2}$).  From Eq.~\eqref{balance}, we derive a simple ordinary differential equation for the frequency as a function of time, and, once it is solved, a further integration yields the phase as a function of frequency. 

The invariant energy $E$ follows from the conservative dynamics of the compact binary at 4PN order, which have been obtained by various groups using different methods: (i) the Arnowitt-Deser-Misner (ADM) Hamiltonian formalism~\cite{JaraS12, JaraS13, DJS14, DJS16} yielded the first derivation of the 4PN energy, although with an ambiguity parameter obtained by matching the near-zone computation to results imported from gravitational self-force (GSF)~\cite{BiniD13}; (ii) the Fokker Lagrangian formalism in harmonic coordinates~\cite{BBBFMa, BBBFMb,BBBFMc,BBFM17} derived the complete result, without ambiguity and without resorting to GSF, by using a specific regularization procedure which was proven to be equivalent to dimensional regularization; (iii) the effective field theory (EFT) approach~\cite{FS4PN,FStail,GLPR16,FMSS16,PR17,FS19,FPRS19,Blumlein20} rederived the 4PN energy by using dimensional regularization. From this series of works, the binary's invariant energy was obtained as the Noetherian quantity associated with temporal translation, and reads at 4PN order (see \emph{e.g.}~\cite{Blumlein21,Bini:2020nsb} for partial results up to 6PN order):
\begin{align}\label{Ex}
	E &= -\frac{m \nu c^2 x}{2} \Biggl\{ 1 + \biggl( - \frac{3}{4} -
	\frac{\nu}{12} \biggr) x + \biggl( - \frac{27}{8} +
	\frac{19}{8} \nu - \frac{\nu^2}{24} \biggr) x^2 \nonumber\\
	&\qquad\quad + \biggl[ - \frac{675}{64} + \biggl(
	\frac{34445}{576} - \frac{205}{96} \pi^2 \biggr) \nu -
	\frac{155}{96} \nu^2 - \frac{35}{5184} \nu^3 \biggr] x^3
	\nonumber\\ 
&\qquad\quad + \Biggl[ - \frac{3969}{128} +
	\biggl(-\frac{123671}{5760}+\frac{9037}{1536}\pi^2 +
	\frac{896}{15}\gamma_\text{E}+ \frac{448}{15} \ln(16
	x)\biggr)\nu\nonumber\\
& \qquad\quad\qquad +
	\biggl(-\frac{498449}{3456}+\frac{3157}{576}\pi^2\biggr)\nu^2
	+\frac{301}{1728}\nu^3 + \frac{77}{31104}\nu^4\Biggr] x^4 +
	\mathcal{O}\bigl(x^5\bigr)
	\Biggr\} \,.
\end{align}
We denote by $\nu\equiv m_1 m_2/m^2$ the symmetric mass ratio ($\gamma_\text{E}$ is the Euler constant). Since there are no terms of half-integer PN order for circular orbits, this expression is actually valid up to 4.5PN order (as indicated by the final error term).

The second input is the energy flux, which we have computed using the PN-MPM formalism applied to compact binaries at 4.5PN beyond the leading quadrupole formula.
Crucial to this computation was the recently-completed source mass quadrupole moment at 4PN order~\cite{MHLMFB20,MQ4PN_IR,MQ4PN_renorm,MQ4PN_jauge}, the source current quadrupole moment at 3PN order~\cite{HFB_courant} and the non-linear tail-of-memory effect~\cite{TLB22,TB23}. We provide the technical details of the derivation in the companion paper~\cite{BFHLT_II}, and report here only the final result:
\begin{align}\label{Fx}
\mathcal{F} = \frac{32c^5}{5G}\nu^2 x^5 \Biggl\{
&
1 
+ \biggl(-\frac{1247}{336} - \frac{35}{12}\nu \biggr) x 
+ 4\pi x^{3/2}
+ \biggl(-\frac{44711}{9072} +\frac{9271}{504}\nu + \frac{65}{18} \nu^2\biggr) x^2 
+ \biggl(-\frac{8191}{672}-\frac{583}{24}\nu\biggr)\pi x^{5/2}
\nonumber\\
& 
+ \Biggl[\frac{6643739519}{69854400}+ \frac{16}{3}\pi^2-\frac{1712}{105}\gamma_\text{E} - \frac{856}{105} \ln (16\,x) 
+ \biggl(-\frac{134543}{7776} + \frac{41}{48}\pi^2 \biggr)\nu 
- \frac{94403}{3024}\nu^2 
- \frac{775}{324}\nu^3 \Biggr] x^3 
\nonumber\\
&
+ \biggl(-\frac{16285}{504} + \frac{214745}{1728}\nu +\frac{193385}{3024}\nu^2\biggr)\pi x^{7/2} 
\nonumber\\
&
+ \Biggl[ -\frac{323105549467}{3178375200} + \frac{232597}{4410}\gamma_\text{E} - \frac{1369}{126} \pi^2 + \frac{39931}{294}\ln 2 - \frac{47385}{1568}\ln 3 + \frac{232597}{8820}\ln x   
\nonumber\\
& \qquad
+ \biggl( -\frac{1452202403629}{1466942400} + \frac{41478}{245}\gamma_\text{E} - \frac{267127}{4608}\pi^2 + \frac{479062}{2205}\ln 2 + \frac{47385}{392}\ln 3  + \frac{20739}{245}\ln x \biggr)\nu
\nonumber\\
& \qquad
+ \biggl( \frac{1607125}{6804} - \frac{3157}{384}\pi^2 \biggr)\nu^2 + \frac{6875}{504}\nu^3 + \frac{5}{6}\nu^4 \Biggr] x^4
\nonumber\\ 
& 
+ \Biggl[ \frac{265978667519}{745113600} - \frac{6848}{105}\gamma_\text{E} - \frac{3424}{105} \ln (16 \,x)
+ \biggl( \frac{2062241}{22176} + \frac{41}{12}\pi^2 \biggr)\nu
\nonumber\\ 
& \qquad
- \frac{133112905}{290304}\nu^2 - \frac{3719141}{38016}\nu^3 \Biggr] \pi x^{9/2} 
+ \mathcal{O}\bigl(x^5\bigr) \Biggr\}\,.
\end{align}
In the test-mass limit $\nu\to 0$, we exactly retrieve the result of linear black-hole perturbation theory~\cite{Sasa94,TSasa94,TTS96,Fuj14PN,Fuj22PN}. 
Since BH perturbations have recently been extended numerically to second order in the mass ratio $\nu$~\cite{Wardell:2021fyy,Albertini:2022rfe,Albertini:2022dmc}, it would be interesting to verify the consistency of this numerical result with the PN prediction~\eqref{Fx}. Note also that in the case of black holes, the contributions due to the absorption by the BH horizons are not included in the PN calculation, and should be added separately. The BH absorption is a 4PN effect for Schwarzschild black holes~\cite{PS95}, and a 2.5PN effect for spinning ones~\cite{TMT97,Alvi01,Porto:2007qi,Chatz12,Saketh:2022xjb}.

With both~\eqref{Ex} and~\eqref{Fx} in hand, we apply the flux-balance equation~\eqref{balance} and readily obtain the time evolution of the GW frequency. Sophisticated techniques exist to increase the precision of the PN results and the overlap with numerical relativity~\cite{DIS98pade,BuonD99,BIO09,DNorleans}, but we do not discuss them here and simply present the results in the form of a fully expanded Taylor PN series. We employ the time variable
\begin{equation}\label{tauPN}
	\tau \equiv \frac{\nu c^3}{5Gm}\bigl(t_0-t\bigr)\,,
\end{equation}
where $t$ is the coordinate time in the asymptotic radiative coordinate system, and $t_0$ an integration constant. We have the freedom to redefine it as $t_0\longrightarrow t_0 + \alpha \,\frac{G m}{c^3}$ where $\alpha$ is any constant, which amounts to the replacement $\tau \longrightarrow \tau [1 + \alpha \,\nu/(5\tau)]$. Although $t_0$ is not uniquely defined, it might be formally interpreted as the instant of coalescence, when $x\to+\infty$, and then it satisfies $t_0-t= \mathcal{O}(c^5)$ in the PN regime. At the 4PN order, using the fact that $\tau^{-1} = \mathcal{O}(c^{-8})$ is a small 4PN quantity, we conveniently adjust $\alpha$ so as to simplify as much as possible the result:
\begin{align}\label{xtau}
	x &= \frac{\tau^{-1/4}}{4}\Biggl\{ 1 + \biggl( \frac{743}{4032} +
	\frac{11}{48}\nu\biggr)\tau^{-1/4} - \frac{1}{5}\pi\,\tau^{-3/8}
	\nonumber \\ & \quad \quad + \biggl( \frac{19583}{254016} +
	\frac{24401}{193536} \nu + \frac{31}{288} \nu^2 \biggr)
	\tau^{-1/2} + \biggl(-\frac{11891}{53760} +
	\frac{109}{1920}\nu\biggr) \pi \,\tau^{-5/8} 
	\nonumber \\ & \quad
	\quad + \Biggl[-\frac{10052469856691}{6008596070400} +
	\frac{1}{6}\pi^2 + \frac{107}{420}\gamma_\text{E} -
	\frac{107}{3360} \ln\biggl(\frac{\tau}{256}\biggr)
	\nonumber \\ & \quad \quad \qquad +
	\biggl(\frac{3147553127}{780337152} - \frac{451}{3072}\pi^2 \biggr)
	\nu - \frac{15211}{442368}\nu^2 +
	\frac{25565}{331776}\nu^3\Biggr] \tau^{-3/4} 
	\nonumber \\ &
	\quad \quad + \biggl(-\frac{113868647}{433520640} -
	\frac{31821}{143360}\nu + \frac{294941}{3870720}\nu^2\biggr) \pi
	\tau^{-7/8} 
	\nonumber \\ & \quad
	\quad + \Biggl[-\frac{2518977598355703073}{3779358859513036800} + \frac{9203}{215040}\gamma_\text{E} + \frac{9049}{258048}\pi^2 + \frac{14873}{1128960}\ln 2	+ \frac{47385}{1605632}\ln 3 - \frac{9203}{3440640}\ln\tau
	\nonumber \\ & \quad \quad \qquad +
	\Biggl(  \frac{718143266031997}{576825222758400} + \frac{244493}{1128960}\gamma_\text{E} - \frac{65577}{1835008}\pi^2 + \frac{15761}{47040}\ln 2 - \frac{47385}{401408}\ln 3 - \frac{244493}{18063360}\ln\tau\Biggr) \nu 
	\nonumber \\ & \quad \quad \qquad + \Biggl(  - \frac{1502014727}{8323596288} + \frac{2255}{393216}\pi^2 \Biggr)\nu^2 - \frac{258479}{33030144} \nu^3 + \frac{1195}{262144} \nu^4 \Biggr] \tau^{-1}\ln\tau
	\nonumber \\ & \quad
	\quad + \Biggl[-\frac{9965202491753717}{5768252227584000} + \frac{107}{600}\gamma_\text{E} + \frac{23}{600}\pi^2 - \frac{107}{4800}\ln\biggl(\frac{\tau}{256}\biggr) 
	\nonumber \\ & \quad \quad \qquad +
	\biggl( \frac{8248609881163}{2746786775040} - \frac{3157}{30720}\pi^2 \biggr)
	\nu - \frac{3590973803}{20808990720}\nu^2 -
	\frac{520159}{1634992128}\nu^3 \Biggr] \pi \,\tau^{-9/8}
	+ \mathcal{O}\bigl(\tau^{-5/4}\bigr)\Biggr\}\,.
\end{align}
Then, the GW half-phase $\psi$ of the dominant harmonics is related to the binary's orbital phase $\phi$ by
\begin{equation}\label{redshift}
	\psi = \phi - \frac{2\pi G \dM f}{c^3} \ln\biggl(\frac{f}{f_0}\biggr)\,,
\end{equation}
where $\dM$ is the ADM mass of the binary, and $f_0$ is an arbitrary unphysical scale, reflecting the different origins of time between the local coordinates covering the source and the radiative coordinates. The logarithmic phase modulation was determined in~\cite{Wi93,BS93} and is physically due to the scattering of GWs on the Schwarzschild background associated with $\dM$ (\emph{i.e.} GW tails). While the GW half-phase $\psi$ and the corresponding GW frequency $f=\dot{\psi}/\pi$ are directly measurable, the orbital phase $\phi$ can only be inferred \emph{via} the theoretical prediction~\eqref{redshift}. When expressing the results in terms of the GW observables $\psi$ and $f$, the arbitrary scale $f_0$ is canceled out (see Sec. VI B in the detailed paper \cite{BFHLT_II}). The explicit expression of the time-domain GW~half-phase~$\psi(t)$ in terms of $x(t)$ [given by~\eqref{xtau}] reads
\begin{align}\label{phix}
	\psi &= \psi_0 - \frac{x^{-5/2}}{32\nu}\Biggl\{ 1 + \biggl(
	\frac{3715}{1008} + \frac{55}{12}\nu\biggr)x - 10\pi x^{3/2}
	\nonumber \\ & \quad + \biggl( \frac{15293365}{1016064} +
	\frac{27145}{1008} \nu + \frac{3085}{144} \nu^2 \biggr) x^2 +
	\biggl(\frac{38645}{1344} - \frac{65}{16}\nu\biggr) \pi x^{5/2} \ln x
	\nonumber \\ & \quad +
	\Biggl[\frac{12348611926451}{18776862720} - \frac{160}{3}\pi^2 -
	\frac{1712}{21}\gamma_\text{E} - \frac{856}{21} \ln (16\,x)
	\nonumber \\ & \quad \qquad +
	\biggl(-\frac{15737765635}{12192768} + \frac{2255}{48}\pi^2
	\biggr)\nu + \frac{76055}{6912}\nu^2 -
	\frac{127825}{5184}\nu^3\Biggr] x^3 
	\nonumber \\ & \quad +
	\biggl(\frac{77096675}{2032128} + \frac{378515}{12096}\nu -
	\frac{74045}{6048}\nu^2\biggr) \pi x^{7/2}
	\nonumber \\ & \quad +
	\Biggl[ \frac{2550713843998885153}{2214468081745920} - \frac{9203}{126}\gamma_\text{E} - \frac{45245}{756}\pi^2 - \frac{252755}{2646}\ln 2 - \frac{78975}{1568}\ln 3 - \frac{9203}{252}\ln x 
	\nonumber \\ & \quad \qquad +
	\biggl(-\frac{680712846248317}{337983528960} - \frac{488986}{1323}\gamma_\text{E} + \frac{109295}{1792}\pi^2 - \frac{1245514}{1323}\ln 2 + \frac{78975}{392}\ln 3 - \frac{244493}{1323}\ln x \biggr)\nu 
	\nonumber \\ & \quad \qquad \qquad  +
	\biggl( \frac{7510073635}{24385536} - \frac{11275}{1152}\pi^2 \biggr)\nu^2 +
	 \frac{1292395}{96768}\nu^3 - \frac{5975}{768}\nu^4\Biggr] x^{4} 
	\nonumber \\ & \quad +
	\Biggl[ - \frac{93098188434443}{150214901760} + \frac{1712}{21}\gamma_\text{E} + \frac{80}{3}\pi^2 +  \frac{856}{21}\ln (16 x)
	\nonumber \\ & \quad \qquad +
	\biggl( \frac{1492917260735}{1072963584} - \frac{2255}{48}\pi^2\biggr)\nu
	 - \frac{45293335}{1016064}\nu^2 - \frac{10323755}{1596672}\nu^3 \Biggr] \pi x^{9/2} 
	 + \mathcal{O}\bigl(x^5\bigr)\Biggr\}\,,
\end{align}
where the integration constant $\psi_0$ is determined by initial conditions, \emph{e.g.}, when the wave frequency enters the detector's band. The results hereabove~\eqref{xtau}--\eqref{phix} give the prediction of Einstein's general relativity for the GW frequency and phase chirp of non-spinning compact binaries up to 4.5PN precision. 

Up to now we dealt with the time-domain GW half-phase $\psi(t)$. It is useful (especially for data-analysis purposes) to also control the frequency-domain GW half-phase, which we denote $\Psi(F)$. Its PN expansion is obtained by using the stationary phase approximation (SPA)~\cite{Tichy:1999pv} and reads:
\begin{align}\label{phiSPA}
\Psi_\text{SPA} &= 2\pi F\,T_0 + \Psi_0 \nonumber\\
& \nonumber
+\frac{3\,v^{-5}}{128\nu}\Biggl\{
1
+ \biggl(\frac{3715}{756} + \frac{55}{9}\nu\biggr)v^2
- 16\pi v^3 
\\
& \quad\qquad\quad\nonumber
+ \biggl(\frac{15293365}{508032} +	\frac{27145}{504} \nu + \frac{3085}{72} \nu^2 \biggr) v^4 
+ \biggl(\frac{38645}{252} - \frac{65}{3}\nu\biggr) \pi v^5 \ln v
\\
& \quad\qquad\quad\nonumber
+ \Biggl[\frac{11583231236531}{4694215680}- \frac{640}{3}\pi^2- \frac{6848}{21}\gamma_\text{E}- \frac{6848}{21}\ln(4v) + \biggl(-\frac{15737765635}{3048192}+\frac{2255}{12}\pi^2 \biggr)\nu\\
& 
\quad\qquad\qquad\quad\nonumber
+ \frac{76055}{1728}\nu^2- \frac{127825}{1296}\nu^3\Biggr]v^6
\\
& \quad\qquad\quad\nonumber
+ \Biggl[\frac{77096675}{254016} + \frac{378515}{1512}\nu-\frac{74045}{756}\nu^2\Biggr] \pi v^7
\\
& \quad\qquad\quad\nonumber
+ \Biggl[ - \frac{2550713843998885153}{276808510218240}+ \frac{90490}{189}\pi^2 + \frac{36812}{63} \gamma_\text{E} + \frac{1011020}{1323}\ln 2 + \frac{78975}{196} \ln 3 + \frac{18406}{63} \ln v 
\\
& \quad\qquad\qquad\quad\nonumber
+\biggl(\frac{680712846248317}{42247941120} - \frac{109295}{224}\pi^2 + \frac{3911888}{1323} \gamma_\text{E} + \frac{9964112}{1323}\ln 2 - \frac{78975}{49} \ln 3 + \frac{1955944}{1323} \ln v \biggr)\nu
\\
& \quad\qquad\qquad\quad\nonumber
+ \biggl( - \frac{7510073635}{3048192} + \frac{11275}{144} \pi^2\biggr)\nu^2 -\frac{1292395}{12096}\nu^3+\frac{5975}{96}\nu^4 \Biggr] v^8 \ln v
\\
& 
\quad\qquad\quad\nonumber
+ \Biggl[\frac{105344279473163}{18776862720}- \frac{640}{3}\pi^2- \frac{13696}{21}\gamma_\text{E}- \frac{13696}{21}\ln(4v) \\
&
\quad\qquad\qquad\quad 
+ \biggl(-\frac{1492917260735}{134120448}+\frac{2255}{6}\pi^2 \biggr)\nu
+ \frac{45293335}{127008}\nu^2+ \frac{10323755}{199584}\nu^3\Biggr]\pi v^9
+ \mathcal{O}\bigl(v^{10}\bigr)\Biggr\}\,,
\end{align}
where $v\equiv \left(\frac{\pi G m F}{c^3}\right)^{1/3}$ with $F$ being the Fourier frequency, and where $T_0$ and $\Psi_0$ are two integration constants. Again we have adjusted $T_0$ in order to simplify the result (and we have absorbed the usual $-\frac{\pi}{4}$ into $\Psi_0$). The coefficients up to 3.5PN, as well as the 4.5PN piece, are already in use; see \emph{e.g.} App.~A of~\cite{Pratten:2020fqn}.

In order to get intuition on the relative contribution of each PN order to the signal, we provide in Table~\ref{table_pots} rough numerical estimates for the number of accumulated GW cycles in the frequency band of current and future detectors. Our naive estimation does not take the various detector noises into account, and a more realistic estimation should be performed~\cite{AISS05}. Nevertheless, it can be useful to gain insight on the behavior of the PN expansion, which seems to converge well, as we see from Table~\ref{table_pots}. For all the typical compact binaries in Table~\ref{table_pots}, we find that the 4PN and 4.5PN orders amount to about a tenth of a cycle (less than 1 radian). This suggests that systematic errors due to the PN modeling may be dominated by statistical errors and negligible for LISA. However, this should be confirmed by detailed investigations along the lines of~\cite{Owen:2023mid}. 
\begin{table}[h]
	\!\!\!\!\!\!\!\!\centering
	\begin{tabular}{|l||c|c||c|c||c|c|}
		\hline
		Detector & \multicolumn{2}{c||}{LIGO/Virgo} & \multicolumn{2}{c||}{ET}& \multicolumn{2}{c|}{LISA} \\
		\hline
		Masses ($M_\odot$) & $1.4 \times 1.4$ & $ 10\times 10$ &  $1.4 \times 1.4$ & $500 \times 500$ & $10^5 \times 10^5$ & $10^7\times 10^7$ \\
		\hline
		PN order &\multicolumn{6}{|c|}{cumulative number of cycles} \\
		\hline
		\hline
		Newtonian & $2\,562.599$ & $95.502$ & $744\,401.36$ & $37.90$ & $28\,095.39$ & $9.534$ \\
		\hline
		1PN & $143.453$ & $17.879$ & $4\,433.85$ & $9.60$ & $618.31$ & $3.386$ \\
		\hline
		1.5PN & $-94.817$ & $-20.797$ & $-1\,005.78$ & $-12.63$ & $-265.70$ & $-5.181$ \\
		\hline
		2PN & $5.811$ & $2.124$ & $23.94$ &  $1.44$ & $11.35$ & $0.677$ \\
		\hline
		2.5PN & $-8.105$ & $-4.604$ & $-17.01$ & $-3.42$ & $-12.47$ & $-1.821$ \\
		\hline
		3PN & $1.858$ & $1.731$ & $2.69$ & $1.43$ & $2.59$ & $0.876$ \\
		\hline	
		3.5PN & $-0.627$ & $-0.689$ & $-0.93$ & $-0.59$ & $-0.91$ & $-0.383$ \\
		\hline 
		4PN & $-0.107$ & $-0.064$ & $-0.12$ & $-0.04$ & $-0.12$ & $-0.013$ \\
		\hline	
		4.5PN & $0.098$ & $0.118$ & $0.14$ & $0.10$ & $0.14$ & $0.065$ \\
		\hline
	\end{tabular}
	\caption{Contribution of each PN order to the total number of accumulated cycles inside the detector's frequency band, for typical (but non-spinning) quasi-circular compact binaries observed by current and future detectors. We have approximated the frequency bands of LIGO/Virgo, Einstein Telescope (ET) and LISA with step functions, respectively between $\bigl[30\,\text{Hz},10^3\,\text{Hz}\bigr]$, $\bigl[1\,\text{Hz},10^4\,\text{Hz}\bigr]$ and $\bigl[10^{-4}\,\text{Hz},10^{-1}\,\text{Hz}\bigr]$. When the merger occurs within the frequency band of the detector, the exit frequency is taken to be the Schwarzschild ISCO, $f_\text{ISCO}=c^3/(6^{3/2}\pi G m)$. The contributions due to the non-linearities of GR (\emph{e.g.}, tails) increase with the PN order and are detailed in~\cite{BFHLT_II}.}
	\label{table_pots}
\end{table}

Besides the chirp described by the results~\eqref{xtau}--\eqref{phix}, it is also important to compute the wave amplitude, in view of the data analysis of LISA~\cite{HM1,HM2,AISSV07} and high-accuracy comparisons with numerical relativity (see \emph{e.g.}~\cite{Grandclement:2007sb,Boyle:2007ft,Bernuzzi:2011aq,Borhanian:2019kxt}). We decompose the waveform, at leading order in the distance $R$ to the source, onto a basis of spin-weighted spherical harmonics (following the conventions of~\cite{BFIS08,FMBI12})
\begin{equation}\label{modedef}
h_+ - \di h_\times
= \frac{8 G m \nu x}{R c^2} \,\sqrt{\frac{\pi}{5}}\, \sum_{\ell=2}^{+\infty}\sum_{\dm=-\ell}^{\ell} H_{\ell \dm} e^{-\di \dm \psi} Y_{-2}^{\ell \dm}\,,
\end{equation}
where the half-phase variable is given by~\eqref{phix}. All $H_{\ell \dm}$ modes  are currently known at 3.5PN order for spinning, non-precessing, quasi-circular orbits~\cite{ABIQ04,BFIS08,FMBI12,Henry:2022dzx,Henry:2022ccf}. Although we were able to derive the phase with 4.5PN accuracy, the same precision for the modes is yet out of reach, since, even though the 4.5PN radiation-reaction terms in the equations of motion are known~\cite{GII97,Leibovich:2023xpg}, neither the source quadrupole moment nor the non-linear contributions to the GW propagation are fully controlled at 4.5PN order (only the contributions that enter the 4.5PN flux for circular orbits are known). We thus report the extension of the dominant quadrupole mode $(\ell,\dm) = (2,2)$ for non-spinning, quasi-circular orbits up to 4PN order:
\begin{align} \label{h22}
	H_{22} &= 1 + \biggl(-\frac{107}{42}+\frac{55}{42}\nu\biggr) x 
	+2 \pi x^{3/2} 
	+ \biggl(-\frac{2173}{1512}-\frac{1069}{216}\nu+\frac{2047}{1512}\nu^2\biggr)x^2
	+  \left[-\frac{107 \pi }{21} +\left(\frac{34 \pi}{21}-24 \,\di\right)\nu\right]x^{5/2} \nonumber \\
	&\qquad + \Biggl[\frac{27027409}{646800}-\frac{856}{105}\,\gamma_\text{E} +\frac{428\,\di\,\pi }{105}+\frac{2 \pi ^2}{3} + \biggl(-\frac{278185}{33264}+\frac{41 \pi^2}{96}\biggr) \nu -\frac{20261}{2772}\nu^2+\frac{114635}{99792}\nu^3-\frac{428}{105} \ln (16 x)\Biggr]x^3
	\nonumber \\
	&\qquad +  \Biggl[ -\frac{2173\pi}{756} + \biggl(
	-\frac{2495\pi}{378}+\frac{14333\,\di}{162} \biggr)\nu + \biggl(
	\frac{40\pi}{27}-\frac{4066\,\di}{945} \biggr)\nu^2 \Biggr] x^{7/2} 
	\nonumber \\
	&\qquad + \Biggl[- \frac{846557506853}{12713500800} + \frac{45796}{2205}\gamma_\text{E} - \frac{22898}{2205}\di \pi - \frac{107}{63}\pi^2 + \frac{22898}{2205}\ln(16x) \nonumber \\
	&\qquad\qquad  +\biggl(- \frac{336005827477}{4237833600} + \frac{15284}{441}\gamma_\text{E} - \frac{219314}{2205}\di \pi - \frac{9755}{32256}\pi^2 + \frac{7642}{441}\ln(16x)\biggr)\nu \nonumber \\
	&\qquad\qquad +\biggl( \frac{256450291}{7413120} - \frac{1025}{1008}\pi^2 \biggr)\nu^2 - \frac{81579187}{15567552}\nu^3 + \frac{26251249}{31135104}\nu^4 \Biggr] x^4 +
	\mathcal{O}\bigl(x^{9/2}\bigr) \,.
\end{align}
Satisfyingly, this result is in perfect agreement with linear black-hole perturbation theory in the limit when $\nu\to 0$; see App. B of~\cite{TSasa94}. Again, it would be interesting to compare the PN prediction~\eqref{h22} with second-order (numerical or analytical) BH perturbation theory.

\begin{acknowledgments}

We thank Adam Pound for discussions on BH perturbation and GSF results, Bala Iyer for his longstanding support and interest in this project, and Alessandra Buonanno for suggestions. F.L. is grateful to the Institut d'Astrophysique de Paris for its hospitality during this project. He received funding from the European Research Council (ERC) under the European Union’s Horizon 2020 research and innovation program (grant agreement No 817791). G.F. thanks IIT Madras for a Visiting Faculty Fellow position under the IoE program during the completion of this work.

\end{acknowledgments}

\bibliography{ListeRef_phase4PN.bib}

\end{document}